\documentclass[aps,prl,reprint,groupedaddress,showpacs]{revtex4-1}

\usepackage{graphicx}
\usepackage{color} 

\begin{document}


\title{Birth of a subaqueous barchan dune}



\author{Carlos A. Alvarez}
 \email{calvarez@fem.unicamp.br}
\author{Erick M. Franklin}
 \email{franklin@fem.unicamp.br}
 \thanks{Corresponding author}
\affiliation{%
School of Mechanical Engineering, UNICAMP - University of Campinas\\
Rua Mendeleyev, 200, Campinas -- SP, Brazil\\
}%

Accepted manuscript for PHYSICAL REVIEW E 96, 062906 (2017), DOI: 10.1103/PhysRevE.96.062906


\date{\today}

\begin{abstract}
Barchan dunes are crescentic shape dunes with horns pointing downstream. The present paper reports the formation of subaqueous barchan dunes from initially conical heaps in a rectangular channel. Because the most unique feature of a barchan dune is its horns, we associate the timescale for the appearance of horns to the formation of a barchan dune. A granular heap initially conical was placed on the bottom wall of a closed conduit and it was entrained by a water flow in turbulent regime. After a certain time, horns appear and grow, until an equilibrium length is reached. Our results show the existence of the timescales $0.5t_c$ and $2.5t_c$ for the appearance and equilibrium of horns, respectively, where $t_c$ is a characteristic time that scales with the grains diameter, gravity acceleration, densities of the fluid and grains, and shear and threshold velocities.
\end{abstract}


\maketitle


Sand dunes are frequently found in both nature and industry, being present in deserts, rivers \cite{Bagnold_1} and petroleum pipelines \cite{Schaflinger, Stevenson, Allabadidi}, for example. Those forms result from the transport of sand entrained by a fluid flow, and, depending on many factors, such as the flow direction, the strength of the flow, the amount of available sand and the sand cover, different kinds of dunes are observed. When the fluid flow causes moderate shear stress on a granular bed, some grains are displaced by rolling, sliding or by small jumps maintaining contact with the fixed part of the bed. The moving grains form then a mobile granular layer known as bed load. If it takes place over a non-erodible ground or with limited sediment supply under a one-directional flow, they form dunes of crescentic shape, known as barchan dunes \cite{Bagnold_1, Herrmann_Sauermann, Hersen_3}. Under those conditions, barchan dunes are strong attractors that can appear in water and oil pipelines, on the bed of rivers and deserts, and on the surface of other planets \cite{Claudin_Andreotti, Parteli2}, for example. The main feature of a barchan dune, that differentiates it from other bedforms, is its horns pointing downstream. For this reason, the growing of a barchan dune from a granular heap can be related to the appearance and growing of horns.

\begin{figure}[b]
\includegraphics[width=0.6\linewidth]{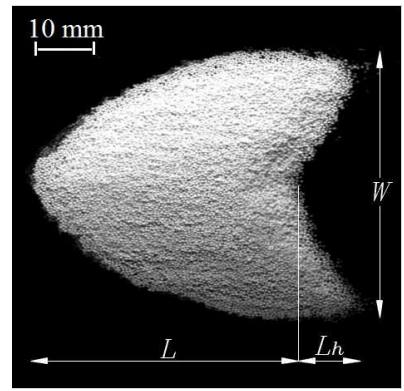}
\caption{Top view of a barchan dune and its characteristic lengths: length $L$, width $W$ and horn length $L_h$.}
	\label{fig:barchan_dimensions}
\end{figure}

Over the past decades, many works were devoted to the equilibrium and minimum size of barchan dunes \cite{Sauermann_1, Hersen_1, Andreotti_1, Hersen_2, Hersen_3, Kroy_B, Parteli, Franklin_8, Kidanemariam}, and to instabilities giving rise to barchans \cite{Kroy_C, Guignier, Parteli3, Khosronejad}. These works investigated the length, width, height and horn scales, the ratio between them, their relation with the saturation length, and the celerity of barchans, for both stable and evolving dunes. The saturation length $L_{sat}$ is the length for the stabilization of sand flux, i.e., it is the length necessary for the sand flux to reach equilibrium conditions with a varying fluid flow \cite{Andreotti_1, Charru_5}. It has been argued that $L_{sat}$ is proportional to an inertial length in the aeolian case \cite{Hersen_1, Sauermann_4}, $L_{drag} = (\rho_s / \rho )d$, and to a relaxation length in the aquatic case \cite{Franklin_8, Charru_5, Charru_3}, $l_d = (u_*/U)d$ or $l_{fall} = (u_*/u_{fall})d$, where $\rho$ is the fluid density, $\rho_s$ is the density of grains, $d$ is the grain diameter, $u_*$ is the shear velocity, $u_{fall}$ is the settling velocity of one grain, and $U$ is the cross-section mean velocity of the fluid. P\"{a}htz et al. \cite{Pahtz_1} derived a theoretical expression for the saturation length that combines $\rho_s/\rho$, $d$ and the fluid velocity. Their expression, which is valid for both aeolian and subaqueous bed load, is in agreement with experimental measurements varying $\rho_s/\rho$ over five orders of magnitude. On the other hand, Claudin and Andreotti \cite{Claudin_Andreotti} showed that the wavelength of dunes scales with $L_{drag}$ over five decades. Therefore, although the fluid flow conditions affect $L_{sat}$, we consider in this paper that $L_{sat}$ is proportional to $L_{drag}$.

Among the studies devoted to the formation of a barchan dune from an initial heap or flat bed, few investigated the behavior of the horns. One of them is Ref. \cite{Khosronejad}, that numerically investigated transverse waves propagating over the horns of already existing barchan dunes and giving rise to new barchans, and showed that the amplitude and wavelength of transverse waves and generated barchans are related. However, none of them investigated the time evolution of growing horns from an initially conical heap. The evolution of the horns for a single barchan dune from an initial conical pile can shed some light on the timescales for the growth of individual barchans, and then characterize the instant corresponding to the appearance of a barchan dune. This question, that has yet to be fully understood, is important to improve our knowledge on the formation of barchans under more complex scenarios, such as sand bars, large sand fields, sand fields over complex topographies, etc.

\begin{figure}[b]
\includegraphics[width=0.9\linewidth]{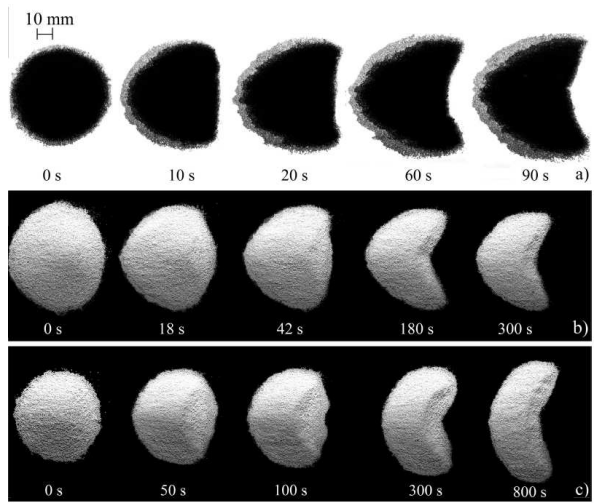}
\caption{Top views of initially conical heaps deformed by the water flow at different times, the water flow is from left to right. a) $Re$ = $1.47 \cdot 10^4 $, $\rho_s$ = 2500 kg/m$^3$ and  $0.40$ mm $\leq\,d\,\leq$ $0.60$ mm. b) $Re$ = $1.67 \cdot 10^4 $, $\rho_s$ = 4100 kg/m$^3$ and  $0.40$ mm $\leq\,d\,\leq$ $0.60$ mm. c) $Re$ = $1.47 \cdot 10^4 $, $\rho_s$ = 4100 kg/m$^3$ and  $0.40$ mm $\leq\,d\,\leq$ $0.60$ mm.}
	\label{fig:barchan_formation}
\end{figure}

In this paper, we present experimental results on the formation of subaqueous barchan dunes from conical heaps showing the existence of characteristic times for the growth and equilibrium of horns. The term equilibrium refers to the crescentic shape of barchan dunes, with horns pointing downstream, which are characteristic of barchan dunes. The experimental device consisted of a water reservoir, two centrifugal pumps, a flow straightener, a 5 $m$ long closed-conduit channel, a settling tank, and a return line. The flow straightener was a divergent-convergent nozzle filled with d = 3 $mm$ glass spheres, the function of which was to homogenize the flow profile at the channel inlet. The channel had a rectangular cross-section (width = 160 $mm$ and height 2$\delta$ = 50 $mm$) and was made of transparent material. The channel test section was 1 $m$ long and started 40 hydraulic diameters (3 $m$) downstream of the channel inlet; therefore, the turbulent flow was completely developed in the test section (the layout of the experimental device is shown in the Supplemental Material \cite{Alvarez2017}). The grains were placed in the test section, which was previously filled with water, and they settled forming a conical heap on the channel bottom wall. Next, a water flow was imposed in the channel, and the heap deformed into a barchan dune. With this procedure, each experiment concerns one single isolated dune. The evolution of the dune was recorded with a CCD (charge coupled device) camera placed above the channel and mounted on a traveling system. Figure \ref{fig:barchan_dimensions} presents one example of top view image, where the characteristic lengths of a barchan dune, namely its length $L$, width $W$ and horn length $L_h$, are shown. Because usually barchan dunes are slightly asymmetric, we consider $L_h$ as the average of both horns. The fluid flow was measured with an electromagnetic flow meter and a 2D-PIV (two-dimensional particle image velocimetry) device.

In the tests, the cross-section mean velocities $U$ were 0.234, 0.294 and 0.333 m/s, corresponding to Reynolds numbers based on the channel height $Re=\rho U 2\delta /\mu$ of $1.16 \cdot 10^4$, $1.47 \cdot 10^4 $ and $1.67 \cdot 10^4 $, respectively, where $\mu$ is the dynamic viscosity of the fluid. The employed fluid was tap water at temperatures within 24 and 26 $^o$C, and the employed grains were round glass beads ($\rho_s = 2500$ kg/m$^3$ and bulk density of 1500 kg/m$^3$) with $0.25$ mm $\leq\,d\,\leq$ $0.50$ mm and $0.40$ mm $\leq\,d\,\leq$ $0.60$ mm, angular glass beads with $0.21$ mm $\leq\,d\,\leq$ $0.30$ mm, and round zirconium beads ($\rho_s = 4100$ kg/m$^3$ and bulk density of 2400 kg/m$^3$) with $0.40$ mm $\leq\,d\,\leq$ $0.60$ mm (see Supplemental Material \cite{Alvarez2017} for the images of the Scanning Electron Microscopy of the used beads). The shear velocities on the channel walls were computed from the velocity profiles acquired by the PIV device and were found to follow the Blasius correlation \cite{Schlichting_1}. They correspond to 0.0143, 0.0177 and 0.0195 m/s for the three flow rates employed. The initial heaps were formed with 10.5 g of glass beads, and with 16.5 g and 10.5 g of zirconium beads, corresponding to initial volumes of 7.0, 6.9 and 4.4 cm$^3$, respectively. Table \ref{tab1} presents the parameters of different flow and grains used in the experiments. For each of the conditions listed, between 2 and 4 test runs were performed and are shown next. In this table, two additional parameters are presented, the particle Reynolds Number, $Re_* = \rho u_* d / \mu$, and the Shields number, $\theta = (\rho u_*^2)/((\rho_s - \rho )gd)$, where $g$ is the acceleration of gravity. The particle Reynolds number is the Reynolds number at the grain scale, and the Shields number is the ratio between the entraining force, which is given by the drag caused by the fluid on each grain, and the resisting force, given by the weight of each grain.

\begin{table}[!ht]	
	\begin{center}
		\begin{tabular}{|c|c|c|c|c|c|c|c|}
			\hline
			Condition & $\rho_s/\rho$ & Surface & $d$ & $Re$ & $Re_*$ & $\theta$ & mass \\ 
			$\cdots$ & $\cdots$ & $\cdots$ & mm & $\cdots$ & $\cdots$ & $\cdots$ & g \\\hline
			a & 2.5 & round & 0.25 -- 0.50 & 1.16 $\cdot$ 10$^4$ & 5 & 0.04 & 10.5 \\\hline
			b & 2.5 & round & 0.25 -- 0.50 & 1.47 $\cdot$ 10$^4$ & 7 & 0.06 & 10.5 \\\hline
			c & 2.5 & round & 0.25 -- 0.50 & 1.67 $\cdot$ 10$^4$ & 7 & 0.07 & 10.5 \\\hline
			d & 2.5 & round & 0.40 -- 0.60 & 1.16 $\cdot$ 10$^4$ & 7 & 0.03 & 10.5 \\\hline
			e & 2.5 & round & 0.40 -- 0.60 & 1.47 $\cdot$ 10$^4$ & 9 & 0.04 & 10.5 \\\hline
			f & 2.5 & round & 0.40 -- 0.60 & 1.67 $\cdot$ 10$^4$ & 10 & 0.05 & 10.5 \\\hline
			g & 2.5 & angular & 0.21 -- 0.30 & 1.16 $\cdot$ 10$^4$ & 2 & 0.09 & 10.5 \\\hline
			h & 2.5 & angular & 0.21 -- 0.30 & 1.47 $\cdot$ 10$^4$ & 3 & 0.13 & 10.5 \\\hline
			i & 2.5 & angular & 0.21 -- 0.30 & 1.67 $\cdot$ 10$^4$ & 3 & 0.16 & 10.5 \\\hline
			j & 4.1 & round & 0.40 -- 0.60 & 1.47 $\cdot$ 10$^4$ & 9 & 0.02 & 10.5 \\\hline
			k & 4.1 & round & 0.40 -- 0.60 & 1.67 $\cdot$ 10$^4$ & 10 & 0.03 & 10.5 \\\hline
			l & 4.1 & round & 0.40 -- 0.60 & 1.47 $\cdot$ 10$^4$ & 9 & 0.02 & 16.5 \\\hline
			m & 4.1 & round & 0.40 -- 0.60 & 1.67 $\cdot$ 10$^4$ & 10 & 0.03 & 16.5 \\\hline
		\end{tabular}
	\end{center}
	\caption{Different tested conditions: $\rho_s/\rho$, state of the grains surface, $d$, $Re$, $Re_*$, $\theta$ and the total mass of each heap.}
	\label{tab1}
\end{table}

\begin{figure}[b]
\includegraphics[width=0.85\linewidth]{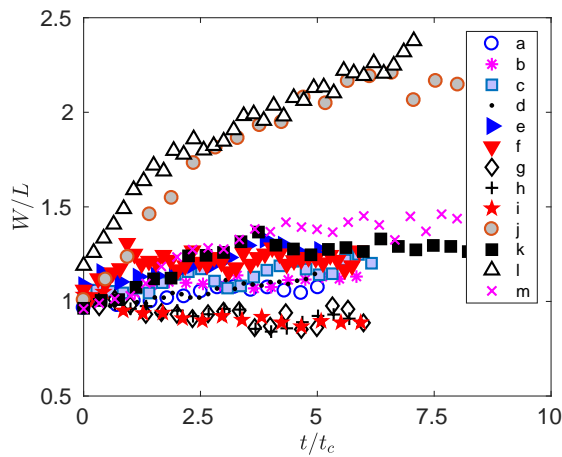}
\caption{$W/L$ versus $t/t_c$ for different flow conditions and solid particles. The cases listed in the key are presented in Tab. \ref{tab1}. This figure shows every $n$ measurements for each experimental condition, where 5 $\leq\,n\,\leq$ 15.}
	\label{fig:WL_time}
\end{figure}

With the conical heap in the channel, the water flow was imposed at a constant rate. Before being displaced over a measurable distance, the pile was deformed and adopted a crescentic shape as shown in Fig. \ref{fig:barchan_formation}. In all cases, a slip face was formed. Although far from the threshold for incipient motion of grains ($\theta \geq 0.03$) barchan dunes have robust ratios between their length $L$, width $W$ and height $h$, both for the aeolian and subaqueous cases \cite{Hersen_1}, we found that the ratio $W/L$ may diverge close to incipient bed load conditions ($0.02 \leq \theta < 0.03$). From Figs. \ref{fig:barchan_formation}b and  \ref{fig:barchan_formation}c, we note the differences in the $W/L$ ratios during the development of barchan dunes under different flow conditions. The larger values of $W/L$ close to threshold conditions persist when barchans have reached equilibrium. As shown in Fig. \ref{fig:WL_time}, which presents the time evolution of $W/L$, the dunes consisting of $0.40$ mm $\leq\,d\,\leq$ $0.60$ mm zirconium beads under $Re$ = $1.47 \cdot 10^4$ (cases j and l of Tab. \ref{tab1}) have $W/L$ values around twice the values presented by dunes farther from the threshold. The cases listed in the key are presented in Tab. \ref{tab1} (see Supplemental Material for $W/L$ versus time for all the test runs). In Fig. \ref{fig:WL_time}, the time was normalized by a characteristic time $t_c$ computed as the length of the bedform divided by its celerity $C$, the latter obtained from the flux rate of grains, $q$, divided by $L_{sat}$ \cite{Claudin_Andreotti}. Because we are interested in subaqueous barchans, we considered $q$ given by the Meyer-Peter and M\"{u}ller equation \cite{Mueller}, and $L_{sat} \propto L_{drag}$. In this manner, $t_c$ is a characteristic time for the displacement of barchans, computed by Eq. \ref{eq:tc},

\begin{equation}
t_c \,=\, \frac{L_{eq}}{C} \,=\, \frac{L_{eq} \left( \rho_s /\rho \right) \left( \rho_s /\rho -1 \right) gd}{\left( u_*^2 - u_{th}^2 \right) ^{3/2}}
\label{eq:tc}
\end{equation}

\noindent where $u_{th}$ is the fluid threshold velocity for incipient motion of grains, $L_{eq}$ is the barchan length at equilibrium, and $g$ is the acceleration of gravity. The threshold velocity was computed in accordance to Ref. \cite{Andreotti_1}. In this way, the commonly reported value $W/L \approx 1$ \cite{Hersen_1, Franklin_8} is not a good indication of the formation and equilibrium of a barchan dune close to threshold conditions (Fig. \ref{fig:barchan_formation}c). As noted by Ref. \cite{Kroy_D}, there is evidence that the scale invariance of barchan dunes is broken in the longitudinal direction and, in order to test similarity and scaling laws, the longitudinal scales shall be addressed. 

\begin{figure}[b]
\begin{center}
	\begin{tabular}{c}
	\includegraphics[width=0.70\columnwidth]{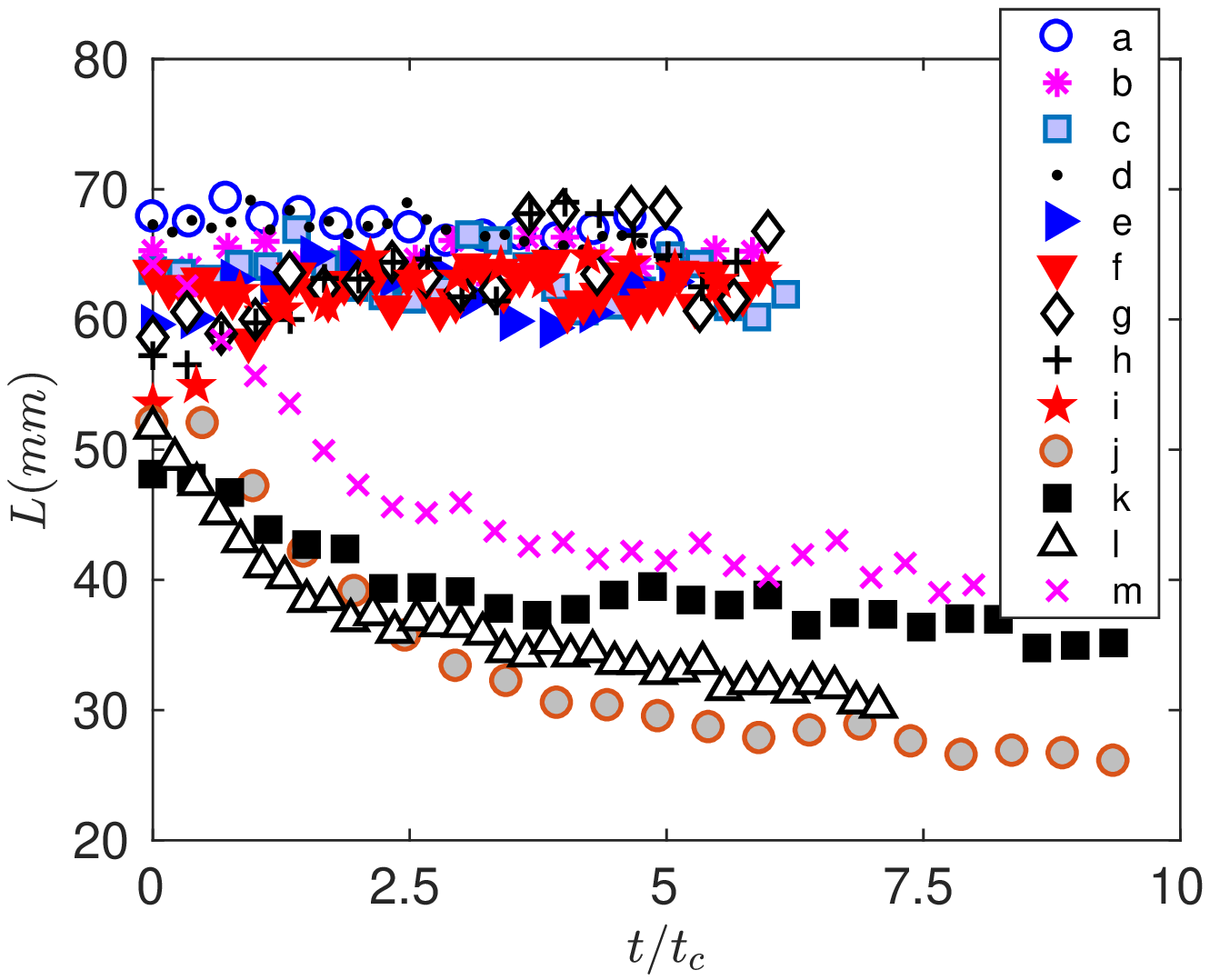}\\
	(a)\\
	\includegraphics[width=0.65\columnwidth]{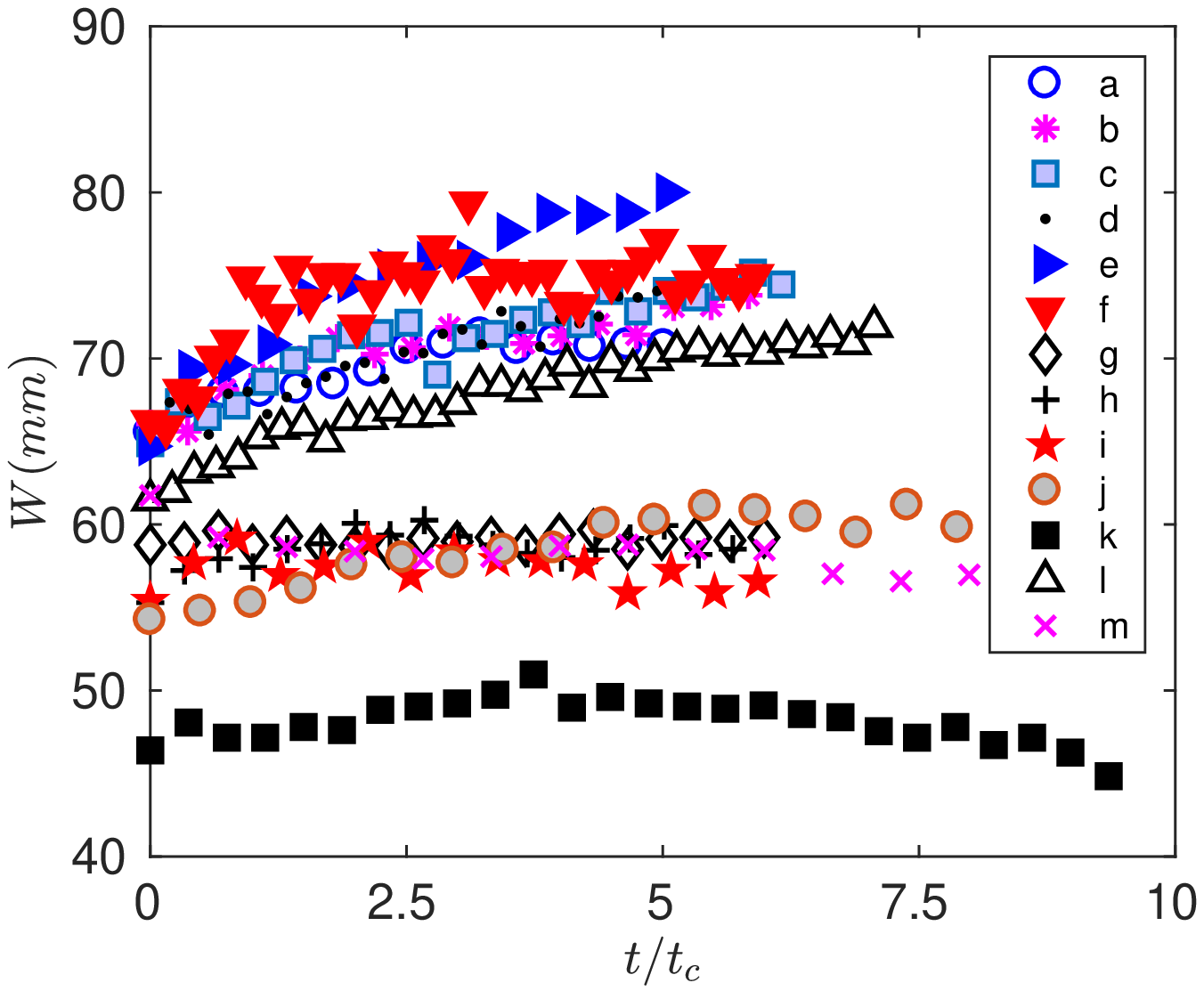}\\
	(b)
	\end{tabular} 
\end{center}
    \caption{(a) $L$ versus $t/t_c$ and (b) $W$ versus $t/t_c$  for different flow conditions and solid particles. The cases listed in the key are presented in Tab. \ref{tab1}. This figure shows every $n$ measurements for each experimental condition, where 5 $\leq\,n\,\leq$ 15.}
    \label{fig:L_time}
\end{figure}

Figures \ref{fig:L_time}a and \ref{fig:L_time}b present the time evolutions of the length and width of dunes, $L$ and $W$ versus $t/t_c$, respectively, for different flow conditions and solid particles. The cases listed in the key are presented in Tab. \ref{tab1} (see Supplemental Material for $L$ and $W$ versus time for all the test runs). From Fig. \ref{fig:L_time}a, we observe different behaviors from the initial piles, with the heaps consisting of heavier zirconium beads decreasing over time while those of glass beads remain constant or increase slightly. The timescales to attain the crescentic shape are very different, being around 2.5 $t_c$ for the glass beads and 6.0 $t_c$ for the zirconium beads. In the case of $W$, Fig. \ref{fig:L_time}b shows that for the different grains and flow conditions the behavior is distinct and a specific timescale cannot be determined.

Although close to threshold $W/L$ diverges and equilibrium times based on $L$ and $W$ are very different, in all cases the conical pile evolves to a crescentic shape with the growing of horns. Therefore, we investigate the time evolution of horns in order to determine when a barchan dune appears and when it reaches equilibrium.

\begin{figure}[b]
\begin{center}
		\includegraphics[width=0.95\columnwidth]{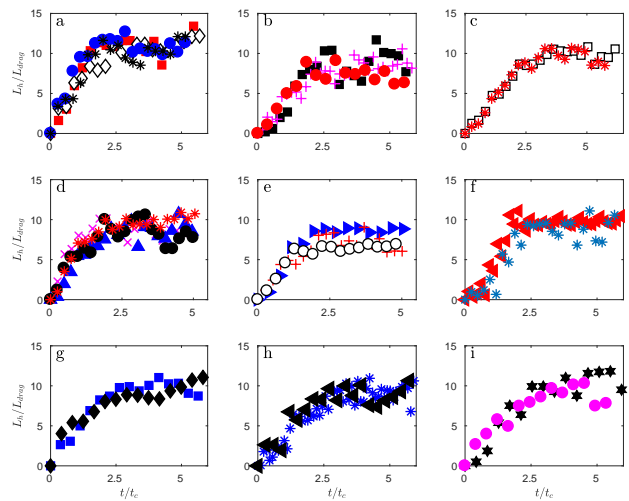}\\	
\end{center}
    \caption{$L_h/L_{drag}$ versus $t/t_c$ for glass beads under different flow conditions over the entire duration of test runs. This figure shows every $n$ measurements for each experimental condition, where 5 $\leq\,n\,\leq$ 15. The cases (a) to (i) are summarized in Tab. \ref{tab1}, and each different symbol in each graphic corresponds to a different test run.}
    \label{fig:horns_glass}
\end{figure}

\begin{figure}[b]
\begin{center}
		\includegraphics[width=0.95\columnwidth]{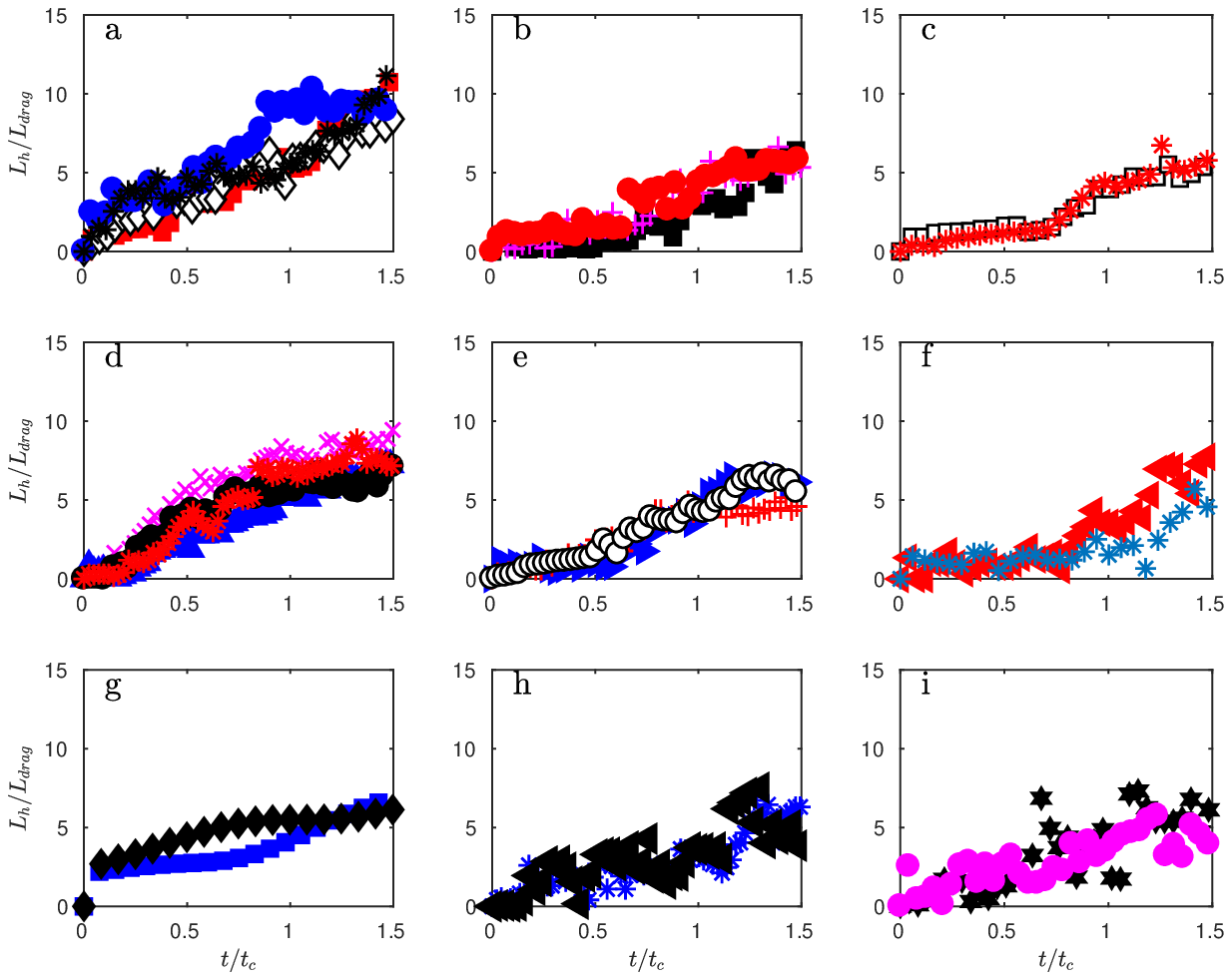}\\	
\end{center}
    \caption{Detail for 0 $\leq \,t/t_c\,\leq$ 1.5 of $L_h/L_{drag}$ versus $t/t_c$ for glass beads under different flow conditions. This figure shows every measurement points within this time interval. The cases (a) to (i) are summarized in Tab. \ref{tab1}, and each different symbol in each graphic corresponds to a different test run.}
    \label{fig:horns_glass_zoom}
\end{figure}

\begin{figure}[b]
\begin{center}
		\includegraphics[width=0.95\columnwidth]{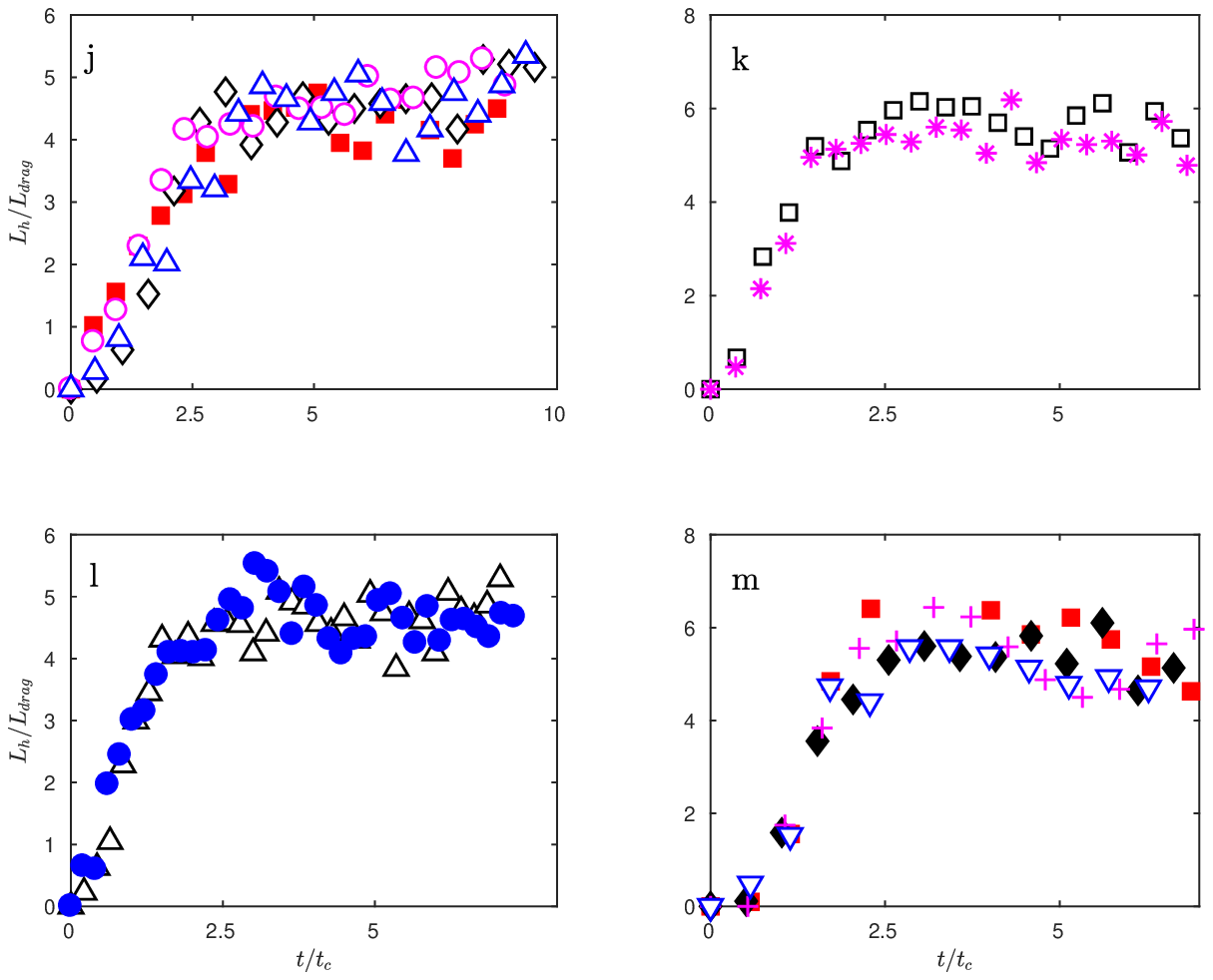}\\	
\end{center}
    \caption{$L_h/L_{drag}$ versus $t/t_c$ for zirconium beads under different flow conditions over the entire duration of test runs. This figure shows every $n$ measurements for each experimental condition, where 5 $\leq\,n\,\leq$ 15. The cases (j) to (m) are summarized in Tab. \ref{tab1}, and each different symbol in each graphic corresponds to a different test run.}
    \label{fig:horns_zirc}
\end{figure}

\begin{figure}[b]
\begin{center}
		\includegraphics[width=0.95\columnwidth]{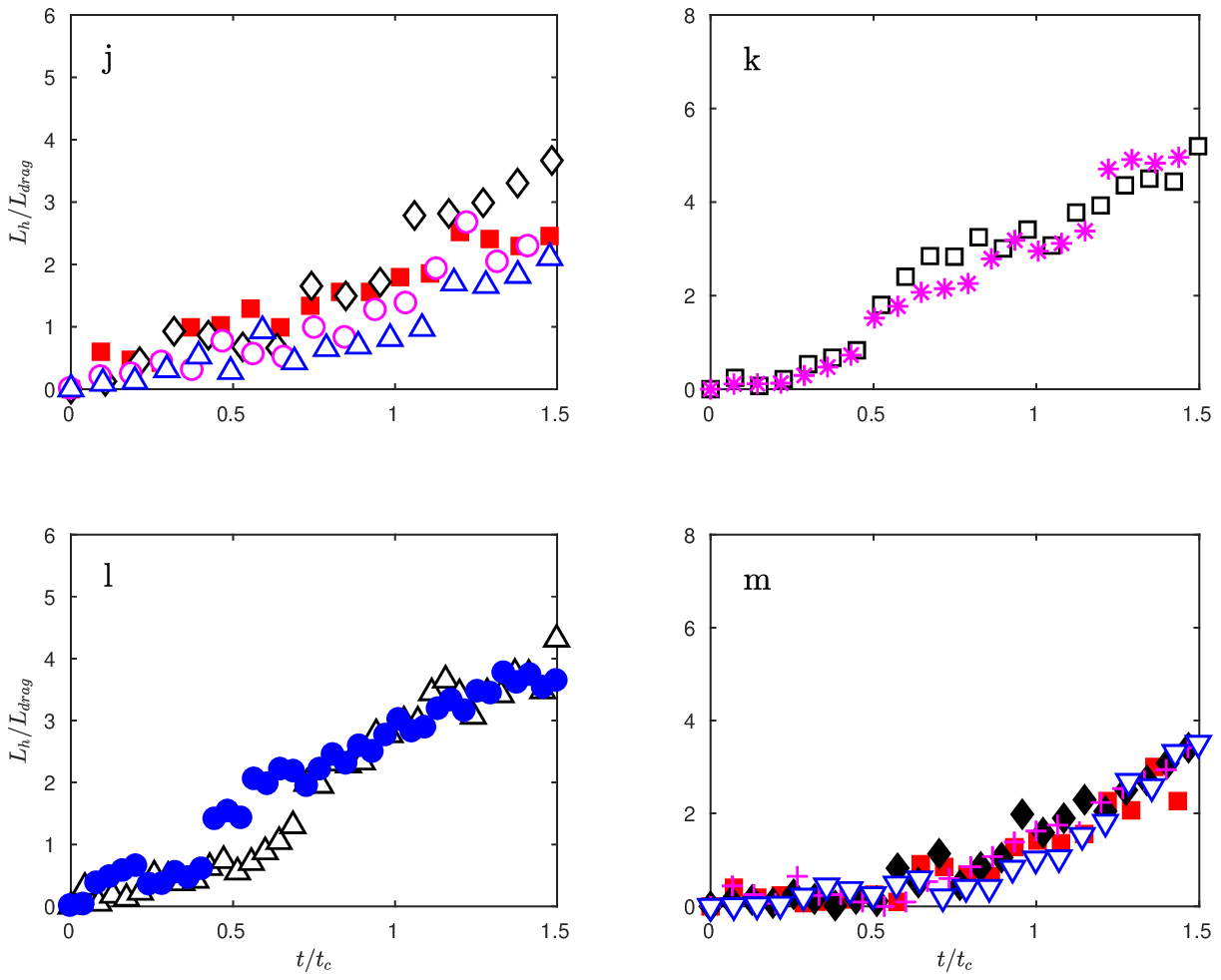}\\	
\end{center}
    \caption{Detail for 0 $\leq \,t/t_c\,\leq$ 1.5 of $L_h/L_{drag}$ versus $t/t_c$ for zirconium beads under different flow conditions. This figure shows every measurement points within this time interval. The cases (j) to (m) are summarized in Tab. \ref{tab1}, and each different symbol in each graphic corresponds to a different test run.}
    \label{fig:horns_zirc_zoom}
\end{figure}

Figures \ref{fig:horns_glass} to \ref{fig:horns_zirc_zoom} present the time evolution of horns for different flow conditions and solid particles. The cases (a) to (m) are summarized in Tab. \ref{tab1}, and each different symbol in each graphic corresponds to a different test run. Figures \ref{fig:horns_glass} and \ref{fig:horns_zirc} present $L_h/L_{drag}$ versus $t/t_c$ over the entire duration of test runs and Figs. \ref{fig:horns_glass_zoom} and \ref{fig:horns_zirc_zoom} present $L_h/L_{drag}$ versus $t/t_c$ from the beginning of test runs until 1.5 $t_c$. By using as characteristic length $L_{drag}$, $L_h/L_{drag} \geq 1$ is the minimum condition to search for the presence of a horn. For the present experiments, 2.5$d$ $\leq\, L_{drag}\, \leq$ 4.1$d$, and, by considering a particle fraction of 50\%, $L_h/L_{drag} \geq 1$ corresponds to structures consisting of 15 to 66 grains as the minimum structures forming a horn. In that case, the appearance of horns, and then the birth of a barchan from the initially conical heap, occurs at $t/t_c \approx 0.5$. From this moment, the horns continue to grow until $t/t_c \approx 2.5$, when they reach an equilibrium length. This is the instant when the barchan reaches its characteristic crescentic shape.

In conclusion, the rise and equilibrium of a barchan dune is well described by the growing of its horns. From Figs. \ref{fig:horns_glass} to \ref{fig:horns_zirc_zoom}, we note that the time evolution of horns on a single dune presents a similar behavior in all cases, with the beginning of growth and the equilibrium shape occurring at the normalized times of approximately 0.5 and 2.5, respectively. Therefore, we propose that the characteristic time given by Eq. \ref{eq:tc} is the timescale for the beginning and growing of single barchan dunes. This timescale, obtained for single dunes under controlled conditions, is important to increase our knowledge on the formation of barchan dunes under more complex scenarios such as sand bars, large sand fields, sand fields over complex topographies, etc.

Carlos A. Alvarez is grateful to SENESCYT (grant no. 2013-AR2Q2850) and to CNPq (grant no. 140773/2016-9). Erick M. Franklin is grateful to FAPESP (grant nos. 2012/19562-6 and 2016/13474-9), to CNPq (grant no. 400284/2016-2) and to FAEPEX/UNICAMP (grant no. 2210/17) for the financial support provided.

\bibliography{references}

\end{document}